\documentclass[preprint,showpacs,preprintnumbers,amsmath,amssymb,nofootinbib]{revtex4}
\usepackage{booktabs}
\usepackage{mathrsfs}
\usepackage{epsfig}
\usepackage{graphicx}
\usepackage{dcolumn}
\usepackage{bm}
\usepackage{amsmath}
\usepackage{slashed}       

\let\jnfont=\rm
\def\NPB#1,{{\jnfont Nucl.\ Phys.\ B }{\bf #1},}
\def\PLB#1,{{\jnfont Phys.\ Lett.\ B }{\bf #1},}
\def\EPJC#1,{{\jnfont Eur.\ Phys.\ Jour.\ C }{\bf #1},}
\def\PRD#1,{{\jnfont Phys.\ Rev.\ D }{\bf #1},}
\def\PRL#1,{{\jnfont Phys.\ Rev.\ Lett.\ }{\bf #1},}
\def\MPLA#1,{{\jnfont Mod.\ Phys.\ Lett.\ A }{\bf #1},}
\def\JPG#1,{{\jnfont J.\ Phys.\ G}{\bf #1},}
\def\CTP#1,{{\jnfont Commun.\ Theor.\ Phys.\ }{\bf #1},}
\def\ZPC#1,{{\jnfont Z.\ Phys.\ C }{\bf #1},}
\def\JHEP#1,{{\jnfont JHEP \ }{\bf #1},}
\def\lsim{\raise0.3ex\hbox{$<$\kern-0.75em\raise-1.1ex\hbox{$\sim$}}}
\def\gsim{\raise0.3ex\hbox{$>$\kern-0.75em\raise-1.1ex\hbox{$\sim$}}}

\begin{document}

\title{ Probing natural SUSY from stop pair production at the LHC}

\author{Junjie Cao$^{1,2}$, Chengcheng Han$^3$, Lei Wu$^3$, Jin Min Yang$^3$,
        Yang Zhang$^{1,3}$
        \\~ \vspace*{-0.3cm} }
\affiliation{ $^1$ Physics Department, Henan Normal University,
     Xinxiang 453007, China\\
       $^2$ Center for High Energy Physics, Peking University,
       Beijing 100871, China \\
      $^3$ Institute of Theoretical Physics,
     Academia Sinica, Beijing 100190, China
     \vspace*{1.5cm}}

\begin{abstract}
We consider the natural supersymmetry scenario in the framework of
the $R$-parity conserving minimal supersymmetric standard model
(called natural MSSM) and examine the observability of stop pair
production at the LHC. We first scan the parameters of this scenario
under various experimental constraints, including the SM-like Higgs
boson mass, the indirect limits from precision electroweak data and
B-decays. Then in the allowed parameter space we study the stop pair
production at the LHC followed by the stop decay into a top quark
plus a lightest neutralino or into a bottom quark plus a chargino.
From detailed Monte Carlo simulations of the signals and
backgrounds, we find the two decay modes are complementary to each
other in probing the stop pair production, and the LHC with
$\sqrt{s}= 14$ TeV and 100 $fb^{-1}$ luminosity is capable of
discovering the stop predicted in natural MSSM up to 450 GeV. If no
excess events were observed at the LHC, the 95\% C.L. exclusion
limits of the stop masses can reach around 537 GeV.

\end{abstract}
\maketitle

\section{INTRODUCTION}

Although the standard model (SM) has been successful in describing
the existing experimental data, it is suffering from the hierarchy
problem and new physics based on certain symmetry is widely expected
to appear at TeV scale to stabilize the electroweak hierarchy
against radiative corrections. This belief was further strengthened
by the recent discovery of the Higgs boson at the Large Hadron
Collider (LHC) with its mass determined around 125 GeV
\cite{ATLAS-CMS-1112}. This mass value agrees well with the
prediction of low energy supersymmetry (SUSY), which is so far the
most promising new physics candidate.

In SUSY, all known bosons and fermions have their supersymmetric
partners, and  the scalar top quarks (called stop $\tilde{t}_i$ with
$i=1,2$), as the top quark partners, can modify the property of the
SM Higgs boson by exactly canceling out the dangerous quadratic
divergence of the top quark loop. Obviously, the experimental
determination of the stop properties is crucial to unravel the
nature of supersymmetry in protecting the Higgs mass at the weak
scale and thus solving the hierarchy problem. In fact, such
activities have been carried out extensively at the hadronic
colliders such as the LHC and the Tevatron
\cite{stop-lhc,gluino-mediated,stop-tev}, but in contrast with the
strong mass bounds (about 1 TeV) on the gluino and the first
generation squarks \cite{susy-lhc}, a relatively light stop (say
about 300 GeV) can not be excluded. Nevertheless, it should be
mentioned that  the recently measured Higgs boson mass around 125
GeV may give some indications for the stop
sector\cite{higgs-mass,125-higgs-cmssm,cao-125-Higgs}. In the
popular MSSM with moderate $\tan \beta$ and large $m_A$, the Higgs
mass is given by \cite{higgs-mass}
\begin{equation}\label{mh}
 m^2_{h}  \simeq M^2_Z\cos^2 2\beta +
  \frac{3m^4_t}{4\pi^2v^2} \ln\frac{M^2_{S}}{m^2_t} +
\frac{3m^4_t}{4\pi^2v^2}\frac{X^2_t}{M_{S}^2} \left( 1 -
\frac{X^2_t}{12M^2_{S}}\right),
\end{equation}
where $v=174 {\rm ~GeV}$, $X_t \equiv A_t - \mu \cot \beta$ and
$M_{S}$ is the average stop mass scale defined by
$M_{S}=\sqrt{m_{\tilde{t}_1}m_{\tilde{t}_2}}$. This expression
indicates that, for the heavier stop $\tilde{t}_2$ around 1 TeV as
discussed above, the lighter stop $\tilde{t}_1$ must be heavier than
about 200 GeV and $|X_t|$ must be larger than 1.5 TeV in order to
push the Higgs mass up to 125 GeV \cite{cao-125-Higgs}. About these
constraints, one should keep in mind that they are independent of
the decay modes of $\tilde{t}_1$, but on the other hand,  they may
be greatly weakened if there exists additional contribution to the
Higgs mass \cite{cao-125-Higgs,125-higgs-nmssm}.

On the theoretical side, there are good reasons to consider at least
one stop significantly lighter than other squarks with a mass around
several hundred GeV. Firstly, in some popular grand unification
models, supersymmetry breaking is usually assumed to transmit to the
visible sector at a certain high energy scale, and then Yukawa
contributions to the renormalization group evolution tend to reduce
stop masses more than other squark masses. Secondly, the chiral
mixing for certain flavor squarks is proportional to the mass of the
corresponding quark, and is therefore more sizable for stops. Such a
mixing will further reduce the mass of the lighter stop. Thirdly, in
the MSSM the minimization conditions of its Higgs potential imply
\cite{mz}
\begin{eqnarray}
\frac{M^2_{Z}}{2}=\frac{(m^2_{H_d}+\Sigma_{d})-(m^2_{H_u}+
\Sigma_{u})\tan^{2}\beta}{\tan^{2}\beta-1}-\mu^{2},
\label{minimization}
\end{eqnarray}
where $m^2_{H_d}$ and $m^2_{H_u}$ represent the weak scale soft SUSY
breaking masses of the Higgs fields, $\mu$ is the higgsino mass
parameter, $\tan\beta\equiv v_u/v_d$. $\Sigma_{u}$ and $\Sigma_{d}$
arise from the radiative corrections to the Higgs potential and the
dominant contribution to the $\Sigma_{u}$ is given by
\begin{eqnarray}
\Sigma_u \sim \frac{3Y_t^2}{16\pi^2}\times m^{2}_{\tilde{t}_i}
\left( \log\frac{m^{2}_{\tilde{t}_i}}{Q^2}-1\right)\;.
\label{rad-corr}
\end{eqnarray}
These two equations indicate that, if the individual terms on the
right hand side of Eq.~(\ref{minimization}) are comparable in
magnitude so that the observed value of $M_Z$ is obtained without
resorting to large cancelations, the natural values of $\mu $ and
$m_{\tilde{t}_i}$ should be around 100 GeV and several hundred GeV
respectively. Numerically, the requirement of $\Sigma_u <M_Z^2/2$
(or $\Sigma_u <v^2$) leads to $m_{\tilde{t}_{1,2}}$ upper bounded by
about 500 GeV (or 1.5 TeV) \cite{sigma-mstop}. Moreover, we note
that a light stop is also phenomenologically needed by the
electroweak baryogenesis \cite{baryogenesis} and may be welcomed by
the dark matter physics \cite{coan}. In the MSSM, although the
gluino contribute to $m^{2}_{\tilde{t}_i}$ at the one-loop level and
to $m^2_{H_u}$ at two-loop level, the corrections are proportional
to $m^{2}_{\tilde{g}}$ and can be greatly enhanced by the large
gluino mass\cite{natural-susy-RS1}. In order to the keep the
naturalness, we expect $m_{\tilde{g}}$ to be lighter than about 3
TeV for $m_{\tilde{t}}<1.5$ TeV. However, since the current results
of searching for the supersymmetry indicate that a gluino with mass
about 1 TeV can safely avoid the LHC constraints, we require 1 TeV
$<m_{\tilde{g}}<$ 3 TeV in our calculation.

Motivated by the theoretical preference and the results of the LHC
search for SUSY, recently the natural MSSM scenario attracted broad
attention \cite{natural-susy-RS1,
natural-susy-1,natural-susy-2,natural-susy-3,natural-susy-4}, which
focuses on the following parameter space of the MSSM
\cite{natural-susy-RS1,natural-susy-3}:
\begin{itemize}
\item $|\mu | \lesssim  100-200$~GeV and $m_{\tilde{t}_{1,2}} \leq  1-1.5$~TeV as preferred by Eq.(\ref{minimization}) and Eq.(\ref{rad-corr});
\item 1 TeV $\lesssim m_{\tilde{g}} \lesssim$ 3-4~TeV to escape the LHC constraint and at same time to avoid spoiling color symmetry;
while the electroweak-ino masses may still be at sub-TeV scale;
\item $m_A\sim \left|m_{H_d}^2\right|^{1\over 2}\lesssim |\mu |\tan\beta$ as suggested by the relation
 $ m_A^2 \simeq 2\mu^2 + m_{H_u}^2 + m_{H_d}^2 + \Sigma_u + \Sigma_d$ and Eq.(\ref{minimization});
\item $m_{\tilde{q}_{1,2}},\ m_{\tilde{l}_{1,2}} \sim 10-50$ TeV to provide a decoupling solution to the SUSY flavor and CP problems.
\end{itemize}
Since the stop are relatively light and sensitive to probing this
scenario, there have been recently many theoretical studies on the
collider signatures of the light directly produced stop in the
$R$-parity conserving and violating MSSM\cite{stop-th}. For example,
by using the top tagging technique, the sensitivity of stop searches
were studied in the hadronic, semi-leptonic and di-leptonic
channels\cite{top-tagging-1,top-tagging-2,top-tagging-3,top-tagging-4}.
In order to suppress the di-leptonic top backgrounds, the authors in
Ref.\cite{baiy} explore some new kinematic observables developed
from $M_{T_2}$ to improve the sensitivity of the stop searches. For
the small mass splitting between stop and top, it is pointed that
the rapidity difference and spin correlation of the daughter
products from stops decay can be helpful to discover the
signal\cite{hanz}. When the stop mass is close to the lightest
supersymmetric particle mass, the monojet signature from
$\tilde{t}_1\tilde{t}^{*}_{1}+j$ production is expected to be useful
in detecting the stop\cite{drees}. If the stop mass is degenerate
with the sum of the masses of its decay products, the searches based
on missing transverse energy ($E^{miss}_T$ or $\slashed E_{T}$) have
significant reach for stop masses above 175 GeV\cite{met}. When the
$R$-parity is violated, the decay modes of the stop will be very
different from the ones in $R$-parity conserving MSSM, such as stop
decaying to dilepton and trilepton final states\cite{rpv}. We also
noted that the constrains on the light stop in the natural SUSY have
been discussed by using the results from sparticles searches at the
LHC, and indicated that they were mild and can be safely avoided
currently\cite{natural-susy-RS1, natural-susy-3, natural-susy-4,
yan}.

In this work, we investigate the potential of the LHC in probing the
lighter stop $\tilde{t}_1$ predicted by the natural MSSM with
$R$-parity, which is based on some considerations: (i) Most of the
studies of the stop searches have been carried out under some
assumptions at the LHC in a model independent way or in simplified
models. It will be meaningful to explore what might happen in a
realistic model like MSSM under the current available experimental
constraints; (ii) Due to the $R$-parity conservation, there will be
sizeable missing energy appearing in the sparticles productions and
decays, which can be easily identified in the LHC data; (iii) One
interesting phenomenological feature of the natural MSSM with
$R$-parity is that both the lightest neutralino and the lighter
chargino are Higgsino-like, and consequently $\tilde{t}_1$ always
decays dominantly into $t \tilde{\chi}_1^0$ and $b \tilde{\chi}_1^+$
with $\tilde{\chi}_1^+ \to \tilde{\chi}_1^0 W^\ast$, which can
greatly simplify the analysis of the $\tilde{t}_1$ detection at the
LHC.

For this purpose, we first scan the parameter space of the natural
MSSM by considering various constraints in Sec. II. Then in Sec. III
we discuss the observability of $\tilde{t}_1$ through the direct
stop pair production in the allowed parameter space by performing
the Monte Carlo simulations for the channel $pp \to \tilde{t}_1
\tilde{t}^{*}_1 \to t\tilde{\chi}_1^0\bar{t}\tilde{\chi}_1^0$ and
the channel $pp \to \tilde{t}_1 \tilde{t}^{*}_1 \to
b\tilde{\chi}_1^+\bar{b}\tilde{\chi}_1^-$. We will present their
corresponding sensitivities for  8 TeV LHC and for 14 TeV LHC
respectively. Finally in Sec. IV, we summarize the conclusions
obtained in this work.

\begin{figure}[htbp]
\includegraphics[width=3.5in,height=3in]{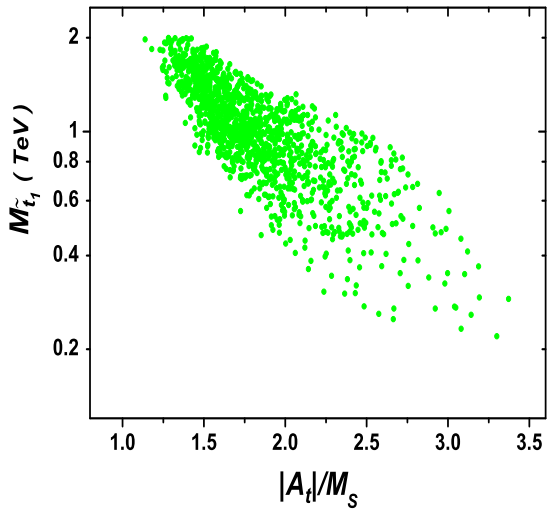}%
\vspace{-0.5cm} \caption{Scatter plot of surviving samples in
natural MSSM, projected on the plane of $m_{\tilde{t}_1}$ versus
$|A_t|/M_S$. All the samples here predict the SM-like Higgs boson in
the mass range $125\pm 2$ GeV. } \label{fig1}
\end{figure}
\section{Scan over the parameter space}

Motivated by the natural MSSM, we scan the parameter space of the
MSSM as follows:
\begin{eqnarray}
&& 1 \le \tan\beta \le 60, \quad 100~\rm GeV \le \mu \le 200~\rm GeV,
\quad 100~\rm GeV \le M_{2}\le 1~\rm TeV, \nonumber \\
&& |A_{t}| \le 3~\rm TeV,
\quad 100~\rm GeV \le (M_{\tilde{Q}_3}, M_{\tilde{U}_3})
\le 2~\rm TeV,
\quad 90~\rm GeV \le M_A \le 1~\rm TeV.
\end{eqnarray}

For other unimportant parameters, we fix all the soft breaking
parameters in the slepton sector and the first two generation sector
at 10 TeV, and we assume $A_{t}=A_{b}$,
$M_{\tilde{U}_3}=M_{\tilde{D}_3}$ and $M_1:M_2 = 1:2$ (inspired by
the grand unification relation).  In our scan, we consider following
constraints:
\begin{figure}[tbp]
\includegraphics[width=3.5in,height=3in]{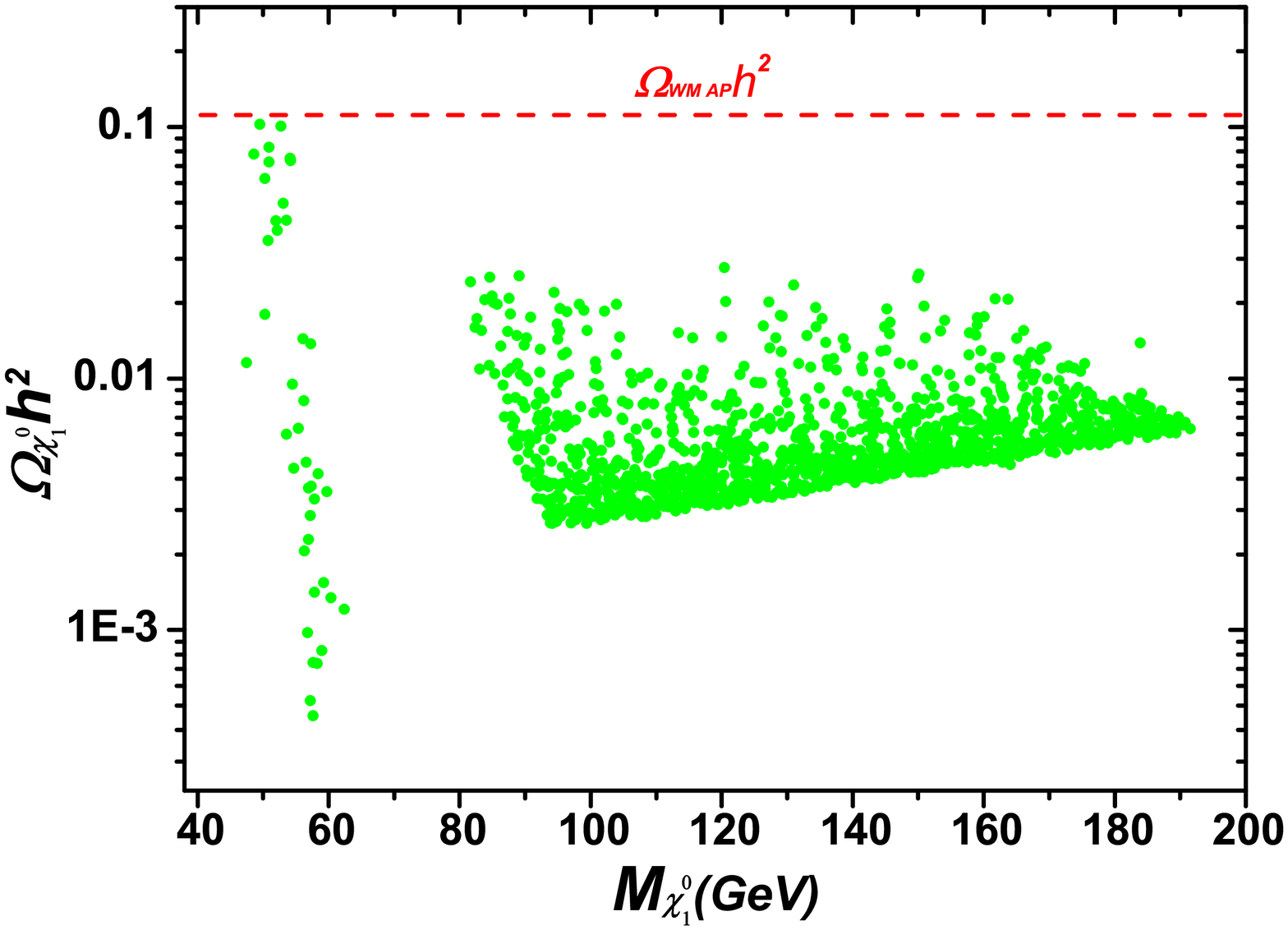}%
\vspace{-0.5cm} \caption{Same as Fig.1, but showing the relic
density of the neutralino dark matter. } \label{fig2}
\end{figure}
\begin{itemize}
\item We require the SM-like Higgs mass within the range $125\pm 2$ GeV. We use
the code \textsf{FeynHiggs2.8.6} \cite{feynhiggs} to calculate the mass and the code
\textsf{HiggsBounds-3.8.0} \cite{higgsbounds} to consider the experimental constraints on
the Higgs sector of the natural MSSM.
\item Since the natural MSSM has important implications in B-physics \cite{natural-susy-b}, we use the code
\textsf{susy$\_$flavor v2.0} \cite{susy-flavor} to consider the constraints from the processes
$B\to X_s \gamma$ and $B_{s(d)}\to \mu^{+} \mu^{-}$.
\item We consider indirect constraints from the precision electroweak observables such as
$\rho_l$, $\sin^2 \theta_{eff}^l$, $m_W$ and $R_b$. We use our own code for such calculation \cite{rb}.
\item We require the thermal relic density of the lightest neutralino (as the dark matter candidate) is below the
WMAP value \cite{wmap}. We use the code \textsf{MicrOmega v2.4} \cite{micromega} to calculate the density.
\end{itemize}

\begin{figure}[tbp]
\includegraphics[width=3.5in,height=3in]{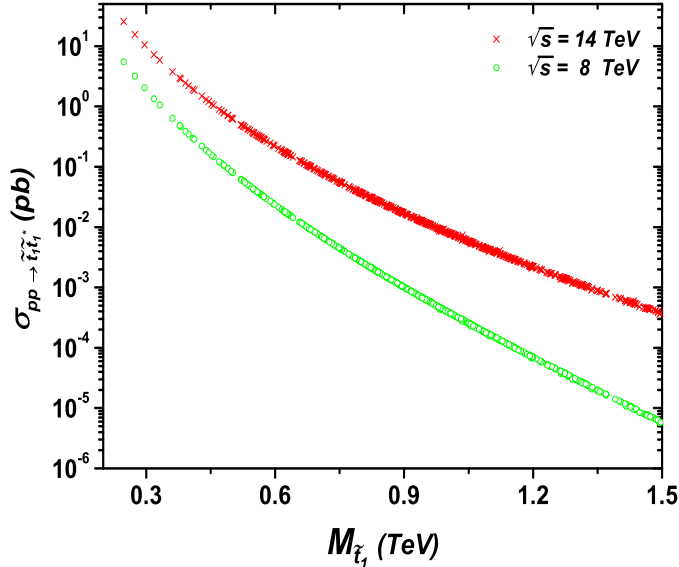}%
\hspace{0.05in}%
\vspace{-0.8cm} \caption{Same as Fig.\ref{fig1}, but showing the
cross section of $\tilde{t}_1\tilde{t}^{*}_{1}$ production versus
the stop mass.} \label{fig3}
\end{figure}
After analyzing the surviving samples, we find they have two
characters. One is that the Higgs mass of $125\pm 2$ GeV requires
$m_{\tilde{t}_1} \geq 220$ GeV and there is a rather strong
correlation between $m_{\tilde{t}_1}$ and the ratio $|A_{t}|/M_S$,
as shown in Fig.\ref{fig1}. Here we further clarify that, if
$M_{\tilde{Q}_3}$ and $M_{\tilde{U}_3}$ are at sub-TeV scale, the
minimum of $m_{\tilde{t}_1}$ will be enhanced to about 300 GeV
\cite{cao-125-Higgs}. The other feature is that for most cases, the
values of $\mu$ are significantly smaller than $M_1$ so that the
lightest neutralino is higgsino-like. Fig.2 indicates that the
surviving samples lie within two isolated regions. We checked that
the lightest neutralino is bino-like in the left region and
higgsino-like in the right region. Here the bino(higgsino)-like
means it is still a mixed state but the dominant component is
bino(higgsino). For the light neutralino dark matter(bino-like), the
main annihilation channel is through exchanging Z boson. The
annihilation cross section is roughly proportional to $1/(4
m_\chi^2-m_Z^2)^2$. When the neutralino mass is about 50GeV$-$60GeV,
the annihilation cross section may be very large, so the relic
density will be less than 0.1. When the neutralino becomes
heavier(60GeV$-$90GeV, neutralino is still bino-like), the
annihilation cross section will drop. The relic density becomes
large and even exceeds the WMAP value, and these samples are
excluded. This is the reason for the gap between 60GeV$-$90GeV. When
the neutralino goes on becoming heavy($>$90 GeV), the dominant
component of the neutralino will be higgsino. The coupling between
neutralino and Higgs gets important and annihilation rate goes up,
then the relic density drops.

About the natural MSSM, we have two comments. One is that in this
scenario the di-photon signal of the SM-like Higgs boson can hardly
be enhanced to satisfy the requirement of the LHC data. This is
because in the framework of the MSSM, there are only two cases which
can enhance the di-photon rate, i.e. the small $\alpha_{eff}$
scenario \cite{small-alpha,diphoton-nmssm} and the light
$\tilde{\tau}$ scenario \cite{higgs-mass,light-stau}, and in each
case a large $\mu$ is needed. In the Ref.\cite{hooper}, the authors
pointed that the light stop with large couplings to Higgs boson in
the SM+\emph{stop} model can improve the SM fitting to the LHC and
Tevatron data by enhancing $\Gamma(h\to \gamma\gamma)$ and
suppressing $\Gamma(h\to gg)$. However, we should note that it does
not mean that the di-photon production rate can reach the
measurement of the LHC in a concrete MSSM model since the reduction
of $\Gamma(h\to gg)$ is usually much stronger than the enhancement
of $\Gamma(h\to \gamma \gamma)$ for large values of
$X_t(X_t=A_t-\mu\tan\beta)$\cite{djouadi}. The other is that
recently the ATLAS collaboration searched for the gluino-mediated
stop pair production followed by the decay $\tilde{t}_{1} \to b
\tilde{\chi}_1^+ \to b \ell \nu \tilde{\chi}_1^0$, which set a lower
bound $m_{\tilde{t}_1} \geq 450 {\rm ~GeV}$ \cite{gluino-mediated}.
This conclusion is not applicable to the our calculations since we
take the gluino mass to be larger than 1TeV in the allowed
parameters space of the natural SUSY.

\section{Observability of stop pair production at the LHC}

In this section we discuss the LHC potential of discovering the stop
through the direct stop pair production in the natural MSSM at
$\sqrt{s}=8,14$ TeV. In Fig.\ref{fig3} we show the $pp \to
\tilde{t}_{1}\tilde{t}^{*}_{1}$ production rate at the next leading
order for the surviving samples. In getting this figure we used the
package \textsf{Prospino2.1} \cite{prospino} and the parton
distribution function CTEQ6.6m \cite{cteq} with the renormalization
scale $\mu_R$ and factorization scale $\mu_F$ setting to
$m_{\tilde{t}_{1}}$. This figure indicates that the maximal values
of the cross section reach 5.5 pb and 25.7 pb for the LHC with
$\sqrt{s}=8$ TeV and $\sqrt{s}=14$ TeV respectively, and with the
increase of the stop mass, the production rates drop rapidly.

\begin{figure}[tbp]
\includegraphics[width=4.0in,height=3in]{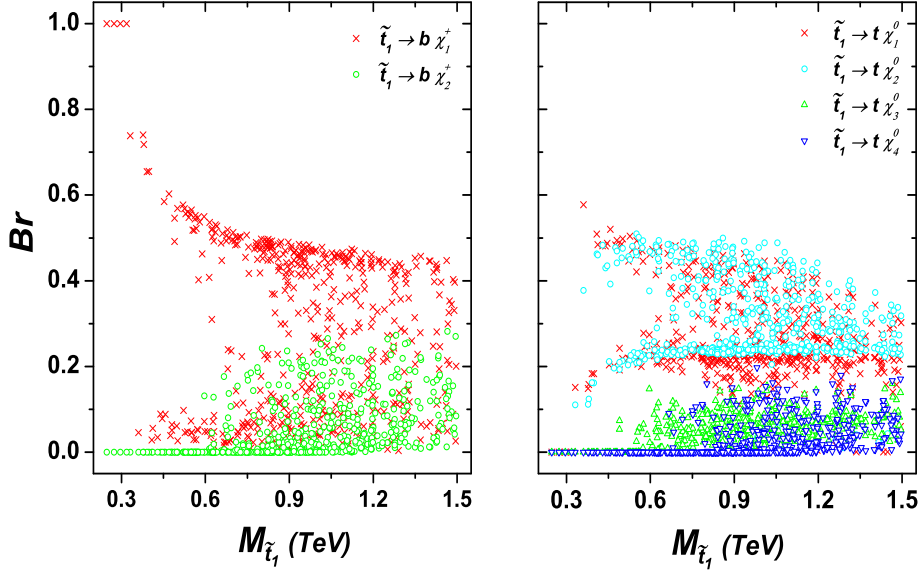}%
\hspace{1.0in}%
\vspace{-0.5cm} \caption{Same as Fig.\ref{fig3}, but showing the
decay branching ratios of the stop.}\label{fig4}
\end{figure}

In Fig.\ref{fig4} we present various decay branching ratios of
$\tilde{t}_{1} $ which are obtained by using the package
\textsf{SDECAY} \cite{sdecay}. This figure indicates that for
$m_{\tilde{t}_1} \leq 320 {\rm ~GeV}$ where the decay channel
$\tilde{t}_{1} \to t\tilde{\chi}_1^0$ does not open up,
$\tilde{t}_{1} $ decays into $b \tilde{\chi}_1^+$ with a ratio of
$100\%$, and as the stop becomes heavier, the branching ratios for
$\tilde{t}_{1} \to b \tilde{\chi}_1^+$ and $\tilde{t}_{1} \to
t\tilde{\chi}_1^0$ may still be around $50\%$. In contrast, the
branching ratios for $\tilde{t}_{1}$ decays into $t
\tilde{\chi}^0_{3,4}$ and $b \tilde{\chi}^+_2$ are usually less than
$20\%$.

In the following we perform detailed Monte Carlo simulations to
investigate the observability of the direct stop pair production at
the LHC. We concentrate on the semi-leptonic analysis with the
b-tagging efficiency 40\%, where the signal is consisted of four
jets(at least one b-jet), one lepton ($e$ or $\mu$), and missing
transverse energy. We first consider the process
\begin{eqnarray}
pp \to \tilde{t}_1 \tilde{t}^{*}_1 \to (t\tilde{\chi}_1^0)
(\bar{t}\tilde{\chi}_1^0) \to
(b\ell^{+}\nu\tilde{\chi}^{0}_{1})(\bar{b}jj\tilde{\chi}^{0}_{1})
~{\rm or}~
(bjj\tilde{\chi}^{0}_{1})(\bar{b}\ell^{-}\bar{\nu}\tilde{\chi}^{0}_{1}).
\label{process1}
\end{eqnarray}
From the ATLAS search for the signal $t\bar{t} + E_T^{miss}$
\cite{stop-lhc}, we can see that the dominant SM background after
the $E_T^{miss}$ and $M_T$ cuts is $t\bar{t}$ di-leptonic channel
with one lost lepton and two additional jets from initial state
radiation to fake the hadronic $W$. Another backgrounds include
$t\bar{t}$ semi-leptonic channel, $t\bar{t}$ di-leptonic channel
with one $\tau$ from top decay misidentified as a jet, $W+$jets and
$t\bar{t}Z$. Here we emphasize that the $t\bar{t}Z$ background
becomes important for a heavy stop and should be considered in
estimating the significance. In our calculation, we normalize the
signal and the $t\bar{t}$ background to their NLO values
\cite{prospino,top-nlo}, and simulate the signal and backgrounds by
\textsf{MadGraph5} \cite{mad5} interfaced with \textsf{PYTHIA}
\cite{pythia} and \textsf{Delphes} \cite{delphes} to carry out the
parton shower and fast detector simulation. We use the anti-$k_t$
algorithm \cite{anti-kt} with the distance parameter $R=0.4$ to
cluster jets and the MLM scheme \cite{mlm} to match our matrix
element with parton shower. We checked that the shapes of the
matched $W+$1,2,3 partons are very similar, and for simplicity, we
take $W+$2 jets samples in our calculations. In our calculations,
since we employ the the variable $M^{W}_{T2}$ defined in
Ref.\cite{baiy}, we checked our results with theirs for the same
parameters at $\sqrt{s}=7$ TeV and found they were consistent with
each other.

\begin{figure}[tbp]
\includegraphics[width=6.5in,height=2.8in]{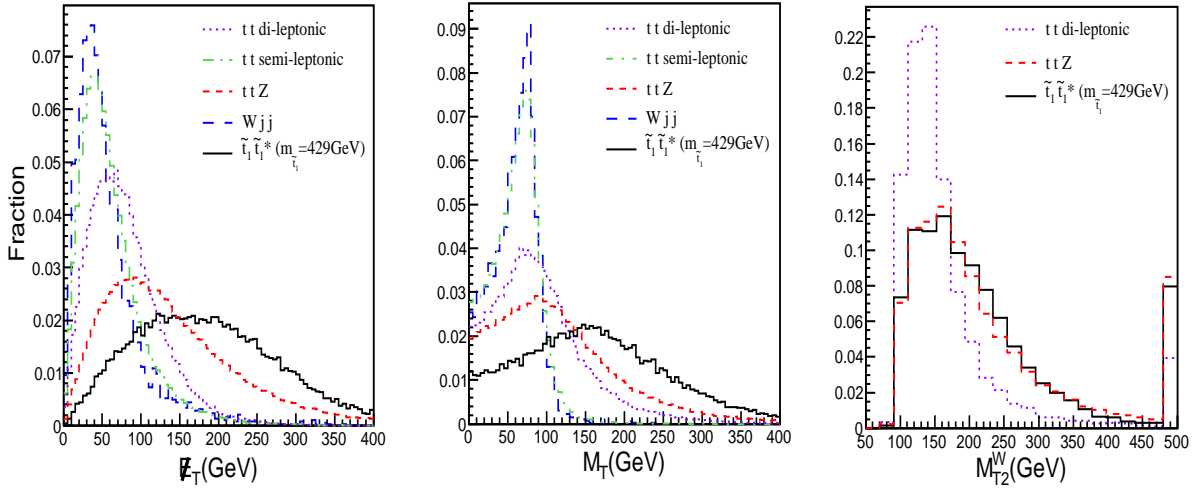}%
\hspace{0.in}%
\vspace*{-1.3cm} \caption{The distributions of
$\ell+$4-jets+$\slashed E_{T}$ with respect to $\slashed E_{T}$,
$M_{T}$ and $M^{W}_{T2}$ for the signal $pp \to \tilde{t}_1
\tilde{t}^{*}_1 \to t\tilde{\chi}_1^0\bar{t}\tilde{\chi}_1^0$ and
backgrounds. In the $M^{W}_{T2}$ distribution, we impose the cuts
$\slashed E_{T}>150$ GeV and $M_{T}>150$ GeV on the events and only
display the di-leptonic $t\bar{t}$, $t\bar{t}Z$ backgrounds and
signal, where the events with wrong or no solution for $M^{W}_{T2}$
are included in the last bin.} \label{fig5}
\end{figure}

In Fig.\ref{fig5}, we show the distributions of $\slashed E_{T}$,
the transverse mass $M_{T}$ defined in \cite{stop-lhc} and
$M^{W}_{T2}$ for the backgrounds and our benchmark point
$m_{\tilde{t}_1}=429$ GeV and $m_{\tilde{\chi}^{0}_1}=110$ GeV with
$\sqrt{s}=8$ TeV (similar results are found for $\sqrt{s}=14$ TeV).
This figure indicates that most events of $W+$jj and semi-leptonic
$t\bar{t}$ backgrounds are characterized by $\slashed E_{T} \leq
100$ GeV and $M_{T} \leq 100$ GeV, and most events of the
di-leptonic $t\bar{t}$ backgrounds are characterized by  $M^{W}_{T2}
\leq 170 {\rm GeV}$, while a significant fraction of the signal may
have larger $\slashed E_{T}$, $M_{T}$ and $M^{W}_{T2}$.
Fig.\ref{fig5} also indicates that the distributions of the
$t\bar{t}Z$ background are quite similar to the signal and are
difficult to be suppressed. Fortunately, the production rate of
$t\bar{t}Z$ is much smaller than the one of
$\tilde{t}_{1}\tilde{t}^{*}_{1}$.

\begin{table}[t] \caption{The significance of stop pair
production  $pp \to \tilde{t}_1 \tilde{t}^{*}_1 \to
t\tilde{\chi}_1^0\bar{t}\tilde{\chi}_1^0$ for 100 $fb^{-1}$
luminosity after imposing various cuts. Here we take
$m_{\tilde{t}_1}=429$ GeV and $\tilde{\chi}^{0}_1=110$ GeV for
illustration. \label{tab1}}
\begin{tabular}{|c|c|c|c|c|c|}
\hline $\slashed E_{T}$-cut (GeV)~&~$M_{T}$-cut (GeV)~&~$M^{W}_{T2}$-cut (GeV)
~&~$S/\sqrt{B}$ (8TeV)~ &~~$S/\sqrt{B}$ (14TeV)\\
\hline ~ ~~150 ~~ & ~~ - ~~   &~~ - ~~   &~~ 1.26 ~~ &~~ 4.05 \\
\hline ~ ~~150 ~~ & ~~ 150 ~~ &~~ - ~~   &~~ 2.75 ~~ &~~ 7.91 \\
\hline ~ ~~150 ~~ & ~~ 150 ~~ &~~ 173 ~~ &~~ 3.11 ~~ &~~ 8.60 \\
\hline
\end{tabular}
\end{table}

In Table I, we present the significance $S/\sqrt{B}$ of our
benchmark point for 100 $fb^{-1}$ luminosity with $\sqrt{s}=8$ TeV
and 14 TeV respectively by sequentially imposing the cuts on
$\slashed E_{T}$, $M_{T}$ and $M^{W}_{T2}$. It can be seen that, for
the given reference point, the cut $M_{T}>150$ GeV can greatly
enhance the significance and $M^{W}_{T2}>173$ GeV further improves
the significance by about $14\%$ for $\sqrt{s}=8$ TeV and $9\%$ for
$\sqrt{s}=14$ TeV to reach 3.11 and 8.60 respectively.

Therefore, for our simulations in the allowed parameters space, we
take the following events selection criteria:

\begin{itemize}
  \item One isolated electron or muon that passes the following requirements;
  \begin{itemize}
    \item Electrons $E_T>25$GeV and $|\eta|<2.47$ without $1.37<|\eta|<1.52$;
    \item Muon:     $E_T>20$GeV and $|\eta|<2.5$;
    \item Events are rejected if they contain a second lepton candidate with $P_T>15$GeV;
  \end{itemize}
  \item Four or more reconstructed jets with $P_T>25GeV$ and $|\eta|<2.5$.
  \item $\slashed E_{T}>$150GeV, $M_{T}>$150GeV, $M^{W}_{T2}>$173GeV.
\end{itemize}
where the basic cuts about $p_T$ and $\eta$ on leptons and jets are
from the ATLAS report\cite{stop-lhc}. In order to improve the signal
sensitivity, we increase the values of ATLAS cut $\slashed E_{T}$
from 100 GeV to 150 GeV to further suppress the semi-leptonic
$t\bar{t}$ background and use the new cut $M^{W}_{T2}>$173 GeV to
reduce the di-leptonic $t\bar{t}$ background in our calculations.

\begin{figure}[htbp]
\includegraphics[width=5in,height=3in]{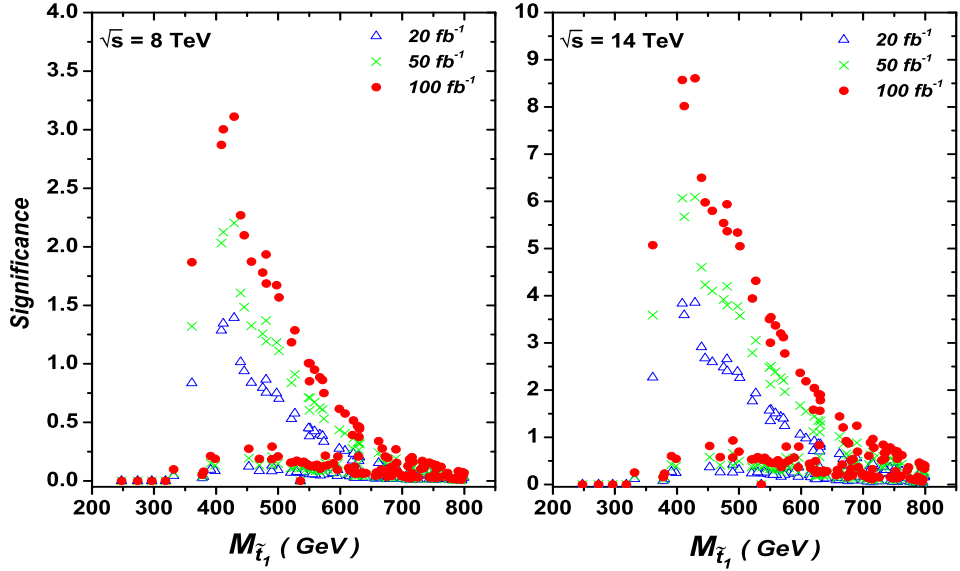}%
\hspace{0.0in}%
\vspace{-0.5cm} \caption{The significance of stop pair production
$pp \to \tilde{t}_1 \tilde{t}^{*}_1 \to
t\tilde{\chi}_1^0\bar{t}\tilde{\chi}_1^0$ for the surviving samples
in the natural MSSM.} \label{fig6}
\end{figure}

In Fig.6 we show the significance of the surviving samples with
$m_{\tilde{t}_{1}}<800$ GeV. This figure indicates that the largest
significance can be reached at $m_{\tilde{t}_{1}}\simeq 430$ GeV
where the significance is about 1.5 for $\sqrt{s}=8$ TeV with 20
$fb^{-1}$ luminosity and 8.5 for $\sqrt{s}=14$ TeV with 100
$fb^{-1}$ luminosity, and with the increase of the stop mass, the
significance drop by one half for $m_{\tilde{t}_{1}}\simeq 500$ GeV
mainly due to the reduction of the production rate.  Our results are
not as optimistic as those in
\cite{top-tagging-1,top-tagging-2,baiy} because we have taken into
account the branch ratio of $\tilde{t}_{1}\to t \tilde{\chi}_1^0$.
Fig.\ref{fig6} also indicates that there are two branches for the
significance in the mass region 320 GeV$<m_{\tilde{t}_1}< $ 600 GeV.
We checked that the upper branch corresponds to high branching ratio
of $\tilde{t}_{1}\to t \tilde{\chi}_1^0$, which varies from $42.1\%$
to $57.7\%$ and results in a large signal rate, while the lower
branch corresponds to a small ratio due to the competition of the
decay mode $\tilde{t}_{1}\to t \tilde{\chi}_1^0$ with $\tilde{t}_{1}
\to b \tilde{\chi}_1^+$ and $\tilde{t}_{1}\to t
\tilde{\chi}^{0}_{2}$.

\begin{figure}[tbp]
\includegraphics[width=6.5in,height=2.8in]{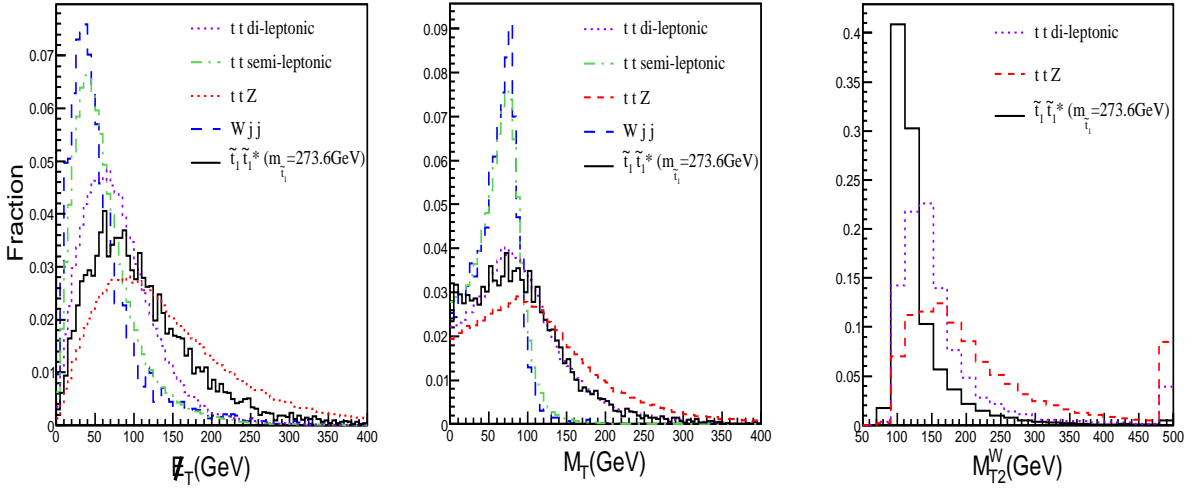}%
\hspace{0.in} \vspace*{-1.3cm} \caption{Same as Fig.\ref{fig5},
except that in plotting $M^{W}_{T2}$ distribution we impose the cuts
$\slashed E_{T}>100$ GeV and $M_{T}>100$ GeV on the events. }
\label{fig7}
\end{figure}

Above analysis implies that, in order to fully explore the parameter
space of the natural MSSM in stop detection,  the decay mode
$\tilde{t}_{1} \to b \tilde{\chi}_1^+$ should also be considered. So
we next consider the process
\begin{eqnarray}
pp \to \tilde{t}_1 \tilde{t}^{*}_1 \to (b\tilde{\chi}_1^+)
(\bar{b}\tilde{\chi}_1^-) \to
(b\ell^{+}\nu\tilde{\chi}^{0}_{1})(\bar{b}jj\tilde{\chi}^{0}_{1})
~{\rm or}~
(bjj\tilde{\chi}^{0}_{1})(\bar{b}\ell^{-}\bar{\nu}\tilde{\chi}^{0}_{1}).
\label{process2}
\end{eqnarray}
Same as in Fig.\ref{fig5}, we show the distributions of the three
variables in Fig.\ref{fig7} for the benchmark point
$m_{\tilde{t}_1}=273.6$ GeV, $m_{\tilde{\chi}^{+}_1}=163.5$ GeV and
$m_{\tilde{\chi}^{0}_1}=156.3$ GeV. Compared with the distribution
in Fig.\ref{fig5}, one can see that more signal events have lower
values of $\slashed E_{T}$ and low $M_{T}$, due to the relatively
light $\tilde{t}_1$. Fig.\ref{fig7} also indicates the $M^{W}_{T2}$
variable is helpless in suppressing the di-leptonic $t\bar{t}$
events and any cut on $M^{W}_{T2}$ may hurt the signal greatly.

\begin{figure}[htbp]
\includegraphics[width=5in,height=3in]{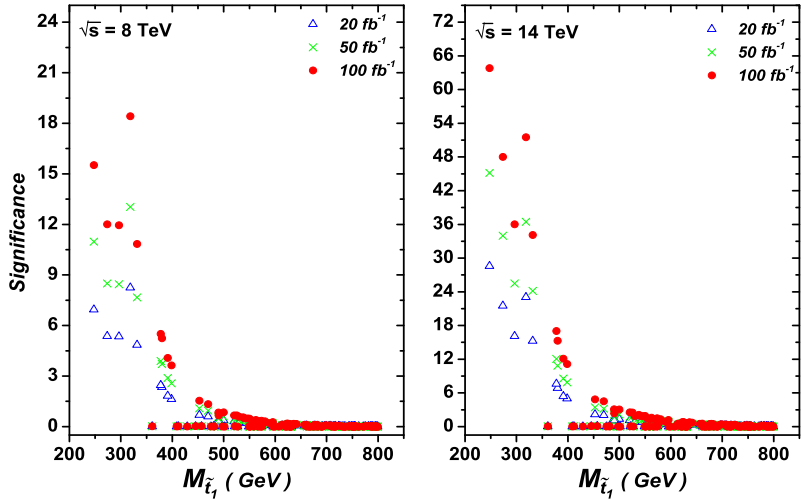}%
\hspace{0.0in}%
\vspace{-0.5cm} \caption{The significance of stop pair production
$pp \to \tilde{t}_1 \tilde{t}^{*}_1 \to (b\tilde{\chi}_1^+)
(\bar{b}\tilde{\chi}_1^-)$ for the surviving samples in the natural
MSSM.} \label{fig8}
\end{figure}

In Fig.\ref{fig8}, we show the significance of the surviving samples
for the process in Eq.(\ref{process2}). In order to keep more signal
events, here we relax the cuts of $\slashed E_{T}$ and $M_{T}$ used
for the process in Eq.(\ref{process1}) as follows:
\begin{eqnarray}
\slashed E_{T}>100~\rm GeV,~~ M_{T}>100~\rm GeV.  \label{cut2}
\end{eqnarray}
From this figure, one can learn that, due to the large stop pair
production rate, the significance for $m_{\tilde{t}_1} = 250$ GeV
may reach 7 for $\sqrt{s}=8$ TeV with 20 $fb^{-1}$ luminosity and 64
for $\sqrt{s}=14$ TeV with 100 $fb^{-1}$ luminosity, but for
$m_{\tilde{t}_1} = 400$ GeV, the maximum value drops to 1.5 and 10
respectively.

\begin{figure}[tbp]
\includegraphics[width=4.5in,height=4in]{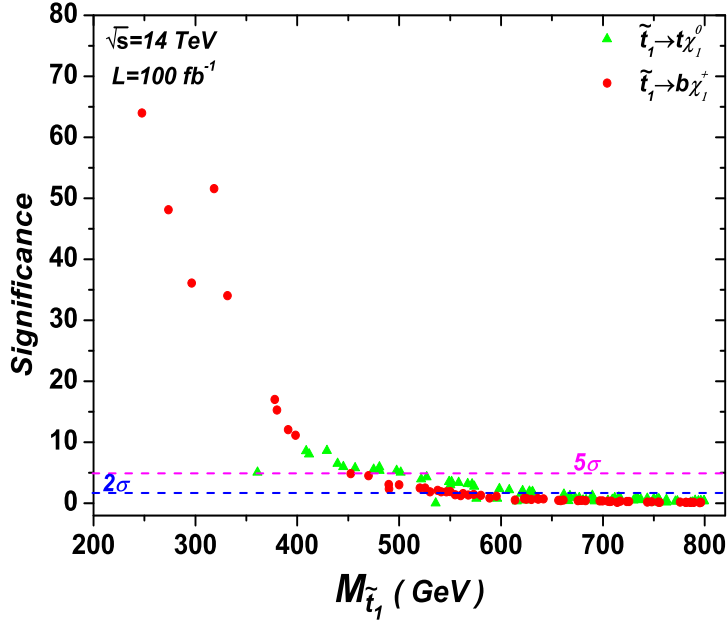}%
\hspace{0.0in}%
\vspace{-0.5cm} \caption{The maximal significance in
Eq.(\ref{process1}) or Eq.(\ref{process2}) as a function of the stop
mass. The bullets (red) are obtained in Eq.(\ref{process2}) while
the triangles (green) are obtained in Eq.(\ref{process1}).}
\label{fig9}
\end{figure}

Finally, we summary the significance of the direct stop pair
production with the above two decay modes of $\tilde{t}_1$ for
$\sqrt{s}=14$GeV and 100 $fb^{-1}$ luminosity. The results are
displayed in Fig.\ref{fig9} where only the maximal significance
under each cut is shown. This figure indicates that, for
$m_{\tilde{t}_1}<$400GeV, detecting the stop pair production through
the chargino decay is more effective, while for $ 400 {\rm GeV} \leq
m_{\tilde{t}_1} \leq 450 {\rm GeV}$ the neutralino decay is more
effective. This figure also indicates that the LHC can discover
$\tilde{t}_1$ predicated in nature MSSM up to $450 {\rm GeV}$. If no
excess events were observed at the LHC, the 95\% C.L. exclusion
limits of the stop masses can go up to around 537 GeV no matter what
decay modes of the stop in the natural MSSM.

\section{conclusion}

In this work we studied the direct stop pair production at the LHC
in the natural MSSM. We scanned over the corresponding parameter space by
considering various experimental constraints and then in the allowed parameter space
we examined the observability of
 the direct stop pair production at the LHC through the semi-leptonic analysis.
We focused on the following two channels
\begin{eqnarray}
pp &\to& \tilde{t}_1 \tilde{t}^{*}_1 \to (t\tilde{\chi}_1^0)
(\bar{t}\tilde{\chi}_1^0) \to
(b\ell^{+}\nu\tilde{\chi}^{0}_{1})(\bar{b}jj\tilde{\chi}^{0}_{1})
~{\rm or}~
(bjj\tilde{\chi}^{0}_{1})(\bar{b}\ell^{-}\bar{\nu}\tilde{\chi}^{0}_{1}), \nonumber \\
pp &\to& \tilde{t}_1 \tilde{t}^{*}_1 \to (b\tilde{\chi}_1^+)
(\bar{b}\tilde{\chi}_1^-) \to
(b\ell^{+}\nu\tilde{\chi}^{0}_{1})(\bar{b}jj\tilde{\chi}^{0}_{1})
~{\rm or}~
(bjj\tilde{\chi}^{0}_{1})(\bar{b}\ell^{-}\bar{\nu}\tilde{\chi}^{0}_{1}),
\nonumber
\end{eqnarray}
and performed detailed Monte Carlo simulations about the signals and
backgrounds. We found that for $m_{\tilde{t}_1} < 400 {\rm GeV}$ the
second channel is better while for $400 {\rm GeV} \leq
m_{\tilde{t}_1} \leq  450 {\rm GeV}$ the first channel is better. We
also found that the LHC with $\sqrt{s}= 14 {\rm TeV}$ and $100
fb^{-1}$ luminosity is capable of discovering $\tilde{t}_1$
predicated in nature MSSM up to $450 {\rm GeV}$. If no excess events
were observed at the LHC, the 95\% C.L. exclusion limits of the stop
masses can reach around 537 GeV in the natural MSSM.

\section*{Note Added}
\begin{figure}[tbp]
\includegraphics[width=4.5in,height=4in]{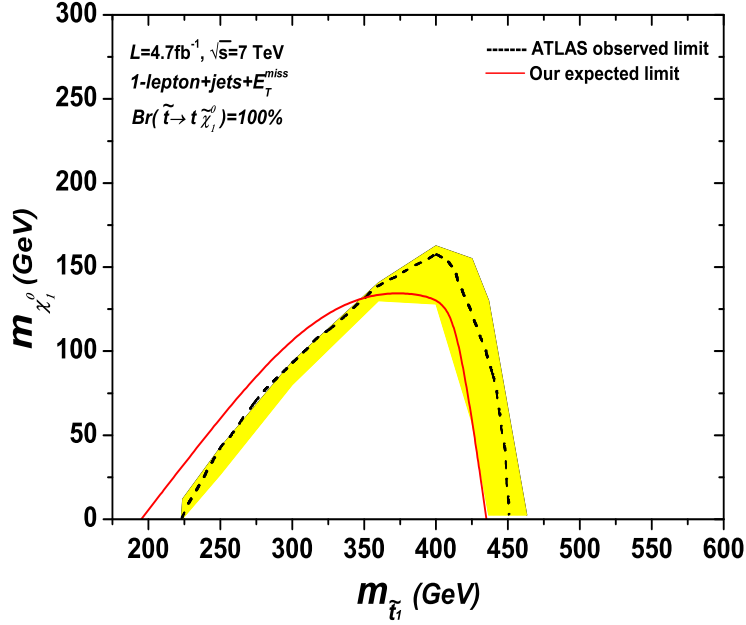}%
\hspace{0.0in}%
\vspace{-0.5cm} \caption{The ATLAS observed and our expected
exclusion limit in the plane of $m_{\tilde{\chi}^{0}_{1}}$ versus
$m_{\tilde{t}_1}$, assuming $BR(\tilde{t}_1 \to t
\tilde{\chi}^{0}_{1})=$100\%.} \label{fig10}
\end{figure}

Very recently, the ATLAS collaboration reported the result of the
direct searching for the stop pair production base on $4.7 fb^{-1}$
of data\cite{atlas-stop}. We validated our simulation by reproducing
the ATLAS exclusion limit according to the assumptions and cuts in
the report as follows:

\begin{itemize}
  \item One isolated electron or muon passing `tight' selection criteria;
  \begin{itemize}
    \item Electrons $E_T>25$GeV and $|\eta|<2.47$;
    \item Muon:     $E_T>20$GeV and $|\eta|<2.4$.
  \end{itemize}
  \item Four or more jets with $|\eta|<2.5$ and $P_T>$80, 60, 40 and 25 GeV,
  and at least one jet to be identified as a b-jet;
  \item $\triangle \phi _{min}>0.8$,where $\triangle \phi _{min}$ is
  the minimum azimuthal separation between the two highest $P_T$ jets
  and the missing transverse momentum direction;
  \item The jet-jet pair having invariant mass $>$ 60 GeV and the smallest
  $\mathrm{\Delta}R$ is selected to form the hadronically decaying $W$ boson.
  The mass $m_{jjj}$ is reconstructed including a third jet closest in $\mathrm{\Delta}R$
  to the hadronic $W$ boson momentum vector and 130 GeV $< m_{jjj}<$ 205 GeV is required;
  \item $E^{miss}_{T}>$150 GeV, $E^{miss}_T/\sqrt{H_T}>$7 GeV$^{1/2}$ and
  $m_{T}>$ 120 GeV;
  \item Events are rejected if they contain additional leptons passing
looser selection criteria. Here we treat the looser selection
criteria as $P_T>$ 15 GeV.
  \item The branch ratio of $\tilde{t}_{1} \to t \tilde{\chi}_1^0$ is
  assumed to be 100\%.
\end{itemize}

In Fig.10, we display the expected exclusion limit from our
simulation. Considering the differences between the fast simulation
and full detector simulation, we can see that our result is
consistent with the ATLAS exclusion limit within the reasonable
error range. We also expect our result can be improved by the
simultaneous fits method used by ATLAS for five signal regions and
three control regions, however, which is beyond the scope of our
simulation. It should be noted that the above stop masses limits can
be avoided in our study, since the stop decays with a mixture of the
branching ratios.

\section*{Acknowledgement}
Lei Wu thanks Xerxes Tata, Zijun Xu and Qiang Li for helpful
discussion about the natural SUSY and MG/ME, and appreciates the
organizers and lecturers at the KIAS school on MadGraph for LHC
physics simulation (Oct. 24-29, 2011, KIAS, Soul). This work was
supported in part by the National Natural Science Foundation of
China (NNSFC) under grant Nos. 10821504, 11135003, 10775039,
11075045, by Specialized Research Fund for the Doctoral Program of
Higher Education with grant No. 20104104110001, and by the Project
of Knowledge Innovation Program (PKIP) of Chinese Academy of
Sciences under grant No. KJCX2.YW.W10.


\begin{thebibliography}{99}
\bibitem{ATLAS-CMS-1112}
  ATLAS Collaboration, ATLAS-CONF-2012-093;
  CMS Collaboration,CMS-PAS-HIG-12-020;
  G.~Aad {\it et al.}  [ATLAS Collaboration],
  Phys.\ Lett.\ B {\bf 710}, 49 (2012);
  S.~Chatrchyan {\it et al.}  [CMS Collaboration],
  Phys.\ Lett.\ B {\bf 710}, 26 (2012).


\bibitem{stop-lhc}
  G.~Aad {\it et al.}  [ATLAS Collaboration],
  Phys.\ Rev.\ Lett.\  {\bf 108}, 041805 (2012).

\bibitem{gluino-mediated}
  G.~Aad {\it et al.}  [ATLAS Collaboration],
  Phys.\ Rev.\ D {\bf 85}, 112006 (2012);
  Phys.\ Rev.\ Lett.\  {\bf 108}, 241802 (2012);
  Phys.\ Rev.\ Lett.\  {\bf 108}, 181802 (2012).

\bibitem{stop-tev}
  V.~M.~Abazov {\it et al.}  [D0 Collaboration],
  Phys.\ Lett.\ B {\bf 710}, 578 (2012);
  Phys.\ Lett.\ B {\bf 696}, 321 (2011);
  T.~Aaltonen {\it et al.}  [CDF Collaboration],
  Phys.\ Rev.\ Lett.\  {\bf 106}, 191801 (2011);
  Phys.\ Rev.\ Lett.\  {\bf 107}, 191803 (2011).

\bibitem{susy-lhc} S. Chatrchyan et al. [CMS collaboration],
                   Phys. Rev. Lett. 107, 221804 (2011);
                   G. Aad et al. [ATLAS collaboration], arXiv:1109.6572 [hep-exp].



\bibitem{higgs-mass}
  M.~Carena, S.~Gori, N.~R.~Shah and C.~E.~M.~Wagner,
  JHEP {\bf 1203}, 014 (2012).

\bibitem{125-higgs-cmssm} see, e.g.,
  J.~Cao {\it et al.}
  Phys.\ Lett.\ B {\bf 710}, 665 (2012);
  H.~Baer, V.~Barger and A.~Mustafayev,
  JHEP {\bf 1205}, 091 (2012);
  L.~Aparicio, D.~G.~Cerdeno and L.~E.~Ibanez,
  JHEP {\bf 1204}, 126 (2012);
  J.~Ellis, K.~A.~Olive and K.~A.~Olive,
  Eur.\ Phys.\ J.\ C {\bf 72}, 2005 (2012);
 C. Balazs et al., arXiv:1205.1568.
A. Fowlie {\it et. al.}, arXiv:1206.0264;


\bibitem{cao-125-Higgs} J.~Cao, {\it et al.},
  JHEP {\bf 1203}, 086 (2012).

\bibitem{125-higgs-nmssm} see, e.g.,
  U.~Ellwanger,
  JHEP {\bf 1203}, 044 (2012);
  J.~F.~Gunion, Y.~Jiang and S.~Kraml,
  Phys.\ Lett.\ B {\bf 710}, 454 (2012);
  U.~Ellwanger and C.~Hugonie,
  arXiv:1203.5048 [hep-ph];
  D. A. Vasquez {\it et. al.},
  Phys.\ Rev.\ D {\bf 86}, 035023 (2012).


\bibitem{mz}
R. Arnowitt and P. Nath, Phys. Rev. D 46, 3981 (1992).

\bibitem{sigma-mstop}
  S.~F.~King, M.~Muhlleitner and R.~Nevzorov,
  Nucl.\ Phys.\ B {\bf 860}, 207 (2012).


\bibitem{baryogenesis}
  P.~Huet and A.~E.~Nelson,
  Phys.\ Rev.\  D {\bf 53} (1996) 4578;
  M.~Carena, G.~Nardini, M.~Quiros and C.~E.~M.~Wagner,
  Nucl.\ Phys.\  B {\bf 812} (2009) 243;
  Y.~Li, S.~Profumo and M.~Ramsey-Musolf,
  Phys.\ Lett.\  B {\bf 673} (2009) 95.

\bibitem{coan}
K.~Griest and D.~Seckel,
  Phys.\ Rev.\  D {\bf 43} (1991) 3191;
   C.~Boehm, A.~Djouadi and M.~Drees, Phys. Rev. {\bf D62} (2000) 035012.

\bibitem{natural-susy-RS1}
  C.~Brust, A.~Katz, S.~Lawrence and R.~Sundrum,
  JHEP {\bf 1203}, 103 (2012).


\bibitem{natural-susy-1}
  D. Feldman, G. Kane, E. Kuflik and R. Lu, Phys. Lett. B 704, 56 (2011);
  H. Baer et al., JHEP 1010, 018 (2010);
  A. Cohen, D. B. Kaplan and A. Nelson, Phys. Lett. B 388, 588 (1996);
  M. Dine, A. Kagan and S. Samuel, Phys. Lett. B 243, 250 (1990).

\bibitem{natural-susy-2}
  J.~L.~Feng and D.~Sanford,
  arXiv:1205.2372 [hep-ph];
  G.~Bhattacharyya and T.~S.~Ray,
  JHEP {\bf 1205}, 022 (2012);
  S.~Krippendorf, H.~P.~Nilles, M.~Ratz and M.~W.~Winkler,
  Phys.\ Lett.\ B {\bf 712}, 87 (2012);
  B.~C.~Allanach and B.~Gripaios,
  JHEP {\bf 1205}, 062 (2012);
  S.~Akula, M.~Liu, P.~Nath and G.~Peim,
  Phys.\ Lett.\ B {\bf 709}, 192 (2012);
  L.~J.~Hall, D.~Pinner and J.~T.~Ruderman,
  JHEP {\bf 1204}, 131 (2012);
  M.~Asano, H.~D.~Kim, R.~Kitano and Y.~Shimizu,
  JHEP {\bf 1012}, 019 (2010);
  R.~Kitano and Y.~Nomura,
  Phys.\ Rev.\ D {\bf 73}, 095004 (2006);
  J.~Hisano, K.~Kurosawa and Y.~Nomura,
  Nucl.\ Phys.\ B {\bf 584}, 3 (2000);
  J.~L.~Feng, K.~T.~Matchev and T.~Moroi,
  Phys.\ Rev.\ D {\bf 61}, 075005 (2000);
  K.~L.~Chan, U.~Chattopadhyay and P.~Nath,
  Phys.\ Rev.\ D {\bf 58}, 096004 (1998);
  G.~W.~Anderson, D.~J.~Castano and A.~Riotto,
  Phys.\ Rev.\ D {\bf 55}, 2950 (1997).


\bibitem{natural-susy-3}
  H.~Baer, V.~Barger, P.~Huang and X.~Tata,
  JHEP {\bf 1205}, 109 (2012).

\bibitem{natural-susy-4}
  M.~Papucci, J.~T.~Ruderman and A.~Weiler,
  arXiv:1110.6926 [hep-ph];

\bibitem{stop-th} see, e.g.,
  A.~Choudhury and A.~Datta,
  JHEP {\bf 1206}, 006 (2012)?
  K.~Huitu, L.~Leinonen and J.~Laamanen,
  Phys.\ Rev.\ D {\bf 84}, 075021 (2011);
  Y.~Kats and D.~Shih,
  JHEP {\bf 1108}, 049 (2011);
  S.~Bornhauser, M.~Drees, S.~Grab and J.~S.~Kim,
  Phys.\ Rev.\ D {\bf 83}, 035008 (2011);
  N.~Bhattacharyya, A.~Choudhury and A.~Datta,
  Phys.\ Rev.\ D {\bf 84}, 095006 (2011);
  D.~Casadei, R.~Konoplich and R.~Djilkibaev,
  Phys.\ Rev.\ D {\bf 82}, 075011 (2010);
  K.~Rolbiecki, J.~Tattersall and G.~Moortgat-Pick,
  Eur.\ Phys.\ J.\ C {\bf 71}, 1517 (2011);
  M.~Perelstein and A.~Weiler,
  JHEP {\bf 0903}, 141 (2009);
  T.~Han, R.~Mahbubani, D.~G.~E.~Walker and L.~-T.~Wang,
  JHEP {\bf 0905}, 117 (2009);
  M.~Carena, A.~Freitas and C.~E.~M.~Wagner,
  JHEP {\bf 0810}, 109 (2008);
  S.~Kraml and A.~R.~Raklev,
  Phys.\ Rev.\ D {\bf 73}, 075002 (2006);
  T. Han at al., Phys. Rev. D {\bf 70}, 055001 (2004).
  A.~Bartl et al.,
  Phys.\ Lett.\ B {\bf 573}, 153 (2003);
  J.~Hisano, K.~Kawagoe and M.~M.~Nojiri,
  Phys.\ Rev.\ D {\bf 68}, 035007 (2003);
  J.~Hisano, K.~Kawagoe, R.~Kitano and M.~M.~Nojiri,
  Phys.\ Rev.\ D {\bf 66}, 115004 (2002);
  J.~M.~Yang and B.~-L.~Young,
  Phys.\ Rev.\ D {\bf 62}, 115002 (2000);


\bibitem{top-tagging-1}
  D.~E.~Kaplan, K.~Rehermann and D.~Stolarski,
  JHEP {\bf 1207}, 119 (2012).

\bibitem{top-tagging-2}
  T.~Plehn, M.~Spannowsky and M.~Takeuchi,
  JHEP {\bf 1208}, 091 (2012);
  Phys. Rev. D 85, 034029 (2012);
  JHEP {\bf 1105}, 135 (2011);
  T. Plehn and M. Spannowsky, arXiv:1112.4441;
  T.~Plehn et al., JHEP {\bf 1010}, 078 (2010);
  T. Plehn, G. P. Salam and M. Spannowsky, Phys. Rev. Lett. 104, 111801 (2010).

\bibitem{top-tagging-3}
  J. Thaler and K. Van Tilburg, JHEP 1202, 093 (2012);
  J. Thaler and K. Van Tilburg, JHEP 1103, 015 (2011);
  J. Thaler and L. -T. Wang, JHEP 0807, 092 (2008).

\bibitem{top-tagging-4}
  K. Rehermann and B. Tweedie, JHEP 1103, 059 (2011);
  M. Jankowiak and A. J. Larkoski, JHEP 1106, 057 (2011);
  L. G. Almeida et al., Phys. Rev. D 82, 054034 (2010); Phys. Rev. D 79, 074012 (2009);
  D. E. Kaplan et al., Phys. Rev. Lett. 101, 142001 (2008).

\bibitem{baiy}
  Y.~Bai, H.~-C.~Cheng, J.~Gallicchio and J.~Gu,
  arXiv:1203.4813 [hep-ph].

\bibitem{hanz}
  Z.~Han, A.~Katz, D.~Krohn and M.~Reece,
  JHEP {\bf 1208}, 083 (2012).


\bibitem{drees}
  M.~Drees, M.~Hanussek and J.~S.~Kim,
  arXiv:1201.5714 [hep-ph];

\bibitem{met}
  D.~S.~M.~Alves et al.,
  arXiv:1205.5805 [hep-ph];

\bibitem{rpv}
  C.~Brust, A.~Katz and R.~Sundrum,
  JHEP {\bf 1208}, 059 (2012);
  H.~-T.~Wei,  et al.,
  JHEP {\bf 1107}, 003 (2011);
  N.~Desai and B.~Mukhopadhyaya,
  JHEP {\bf 1010}, 060 (2010).

\bibitem{yan}
  X.~-J.~Bi, Q.~-S.~Yan and P.~-F.~Yin,
  Phys.\ Rev.\ D {\bf 85}, 035005 (2012);

\bibitem{feynhiggs}
  M.~Frank et al.,
  JHEP {\bf 0702}, 047 (2007);
  G.~Degrassi et al.,
  Eur.\ Phys.\ J.\ C {\bf 28}, 133 (2003);
  S.~Heinemeyer, W.~Hollik and G.~Weiglein,
  Comput.\ Phys.\ Commun.\  {\bf 124}, 76 (2000);
  Eur.\ Phys.\ J.\ C {\bf 9}, 343 (1999).

\bibitem{higgsbounds}
  P.~Bechtle et al.,
  Comput.\ Phys.\ Commun.\  {\bf 182}, 2605 (2011);
  Comput.\ Phys.\ Commun.\  {\bf 181}, 138 (2010).

\bibitem{natural-susy-b}
  K.~Ishiwata, N.~Nagata and N.~Yokozaki,
  Phys.\ Lett.\ B {\bf 710}, 145 (2012).

\bibitem{susy-flavor}
  J.~Rosiek et al.,
  Comput.\ Phys.\ Commun.\  {\bf 181}, 2180 (2010);
  A.~Crivellin, L.~Hofer and J.~Rosiek,
  JHEP {\bf 1107}, 017 (2011).

\bibitem{rb}
  J.~Cao and J.~M.~Yang,
  JHEP {\bf 0812}, 006 (2008).

\bibitem{wmap}
  J. Dunkley {\it et. al.} [WMAP Collaboration], Astrophys. J. Suppl. 180, 306 (2009)


\bibitem{micromega}
  G.~Belanger et al.,
  Comput.\ Phys.\ Commun.\  {\bf 182}, 842 (2011).

\bibitem{small-alpha}
  M.~S.~Carena et al.,
  Eur.\ Phys.\ J.\ C {\bf 26}, 601 (2003).

\bibitem{diphoton-nmssm} J. Cao et al., Phys. Lett. B {\bf 703}, 462 (2011);


\bibitem{light-stau}
  M.~Carena et al.,
  arXiv:1205.5842 [hep-ph].

\bibitem{hooper}
  M.~R.~Buckley and D.~Hooper,
  arXiv:1207.1445 [hep-ph].

\bibitem{djouadi}
  A.~Djouadi,
  Phys.\ Lett.\ B {\bf 435}, 101 (1998);
  [hep-ph/9901237].


\bibitem{prospino}
  W.~Beenakker et al.,
  Nucl.\ Phys.\ B {\bf 515}, 3 (1998);

\bibitem{cteq}
  J.~Pumplin {\it et al.}, JHEP {\bf 0602}, 032 (2006).

\bibitem{sdecay}
  M.~Muhlleitner, A.~Djouadi and Y.~Mambrini,
  Comput.\ Phys.\ Commun.\  {\bf 168}, 46 (2005).

\bibitem{top-nlo}
N. Kidonakis, Phys. Rev. D 82, 114030 (2010);
V. Ahrens et al., Phys. Lett. B 703, 135 (2011);
M. Cacciari et al., Phys. Lett. B 710, 612 (2012);
S. Moch, P. Uwer and A. Vogt, arXiv:1203.6282.

\bibitem{mad5}
  J.~Alwall et al.,
  JHEP {\bf 1106}, 128 (2011).

\bibitem{pythia}
  T.~Sjostrand, S.~Mrenna and P.~Z.~Skands,
  JHEP {\bf 0605}, 026 (2006).

\bibitem{delphes}
  S.~Ovyn, X.~Rouby and V.~Lemaitre,
  arXiv:0903.2225 [hep-ph].

\bibitem{anti-kt}
  M.~Cacciari, G.~P.~Salam and G.~Soyez,
  JHEP {\bf 0804}, 063 (2008).

\bibitem{mlm}
  F.~Caravaglios, M.~L.~Mangano, M.~Moretti and R.~Pittau,
  Nucl.\ Phys.\ B {\bf 539}, 215 (1999).

\bibitem{atlas-stop}
  G.~Aad {\it et al.}  [ATLAS Collaboration],
  arXiv:1208.2590 [hep-ex].


\end{thebibliography}
\end{document}